\documentstyle[aps,twocolumn]{revtex}

\begin{document}

\title{Strangeness Enhancement in the Parton
Model}
\author{Rudolph C. Hwa}
\address{Institute of Theoretical Science and Department of
Physics\\ University of Oregon, Eugene, OR 97403-5203, USA}

\author{C.\ B.\ Yang}
\address{Institute of Particle Physics, Hua-Zhong Normal University,
Wuhan 430079, China}
\date{June 2002} 
\maketitle

\begin{abstract}
Strangeness enhancement in heavy-ion collisions is studied at the
parton level by examining the partition of the new sea quarks
generated by gluon conversion into the strange and non-strange
sectors. The CTEQ parton distribution functions are used as a baseline
for the quiescent sea before gluon conversion. By quark counting
simple constraints are placed on the hadron yields in different
channels. The experimental values of particle ratios are fitted to
determine the strangeness enhancement factor. A quantitative measure
of Pauli blocking is determined. Energy dependence between SPS and
RHIC energies is well described. No thermal equilibrium or
statistical  model is assumed.
\end{abstract}


\section{Introduction}

The production of strange particles in heavy-ion collisions has been
a subject of intense study in the past twenty years, ever since the
proposal that it may reveal a signal of quark-gluon plasma formation
\cite{rm,qm}. Various approaches to the problem have been adopted,
ranging from statistical thermal model \cite{bhs,bck} to simple quark
coalescence model \cite{ab,zbc} to dual parton model
\cite{css}. Despite differences in diverse viewpoints, the major
theme is to explain the phenomenon of strangeness enhancement in
nuclear collisions \cite{qm01}. Although the experimental definition
of strangeness enhancement is the increase of the strangeness content
of the produced hadrons with increasing number of participants from
$pp$ to $AA$ collisions, a more appropriate theoretical description of
strangeness enhancement is in terms of the increase of strange quarks
before hadronization. In this paper we present a quantitative
treatment of the enhancement factor in the framework of the parton
model, and obtain a numerical measure of Pauli blocking.

The point stressed in the original explanation for strangeness
enhancement \cite{rm,qm} is that when the quark degrees of freedom
are liberated, it is easier to create strange quark pairs than
strange hadrons because the $s\bar s$ threshold is lower.
Deconfinement then leads naturally to the possibility of plasma
formation. In our view the quarks have always been the basis for
understanding hadron production even in $pp$ collisions for
$\sqrt s>10$ GeV. The recombination model has been able to reproduce
the low-$p_T$ inclusive distributions in the fragmentation region by
treating the hadronization processes at the parton level
\cite{dh,rch}. Thus the relevance of the quark degrees of freedom in
heavy-ion collisions at SPS is nothing new. In collisions at such
energies the nucleons are broken up and thus deconfined, but it does
not mean that there is thermalized quark-gluon plasma, which no one
would associate with $pp$ collisions.

Once one descends to the parton level, the notion of strangeness
enhancement (SE) can take on a quantitative description in terms of
the strange quark population. The baseline for the unenhanced $s$
quark distribution in the quiescent sea should be pinned down by the
parton distribution functions of the nucleon studied exhaustively by
several groups \cite{ct,mrs,grv}. We shall use the distribution
functions of the CTEQ global analysis
\cite{ct} that fits some 1300 data points obtained for many reactions
in  16 experiments. Their extrapolations to
$Q^2=1$ (GeV/c)$^2$ are presented in the form of graphs available on
the web \cite{cteq4}. Since gluons do not hadronize directly, there
being no glueballs found, they are converted to quark pairs which
subsequently hadronize by recombination. How much the conversion goes
into the strange sector gives us a measure of SE.

Gluon conversion is not a new process that we must consider for
heavy-ion collisions. Even in hadronic collisions gluons must convert
in order to hadronize. Such conversion has been included in the study
of inclusive distributions of hadrons produced in the fragmentation
region in the framework of the valon-recombination model \cite{rch},
and more recently using the CTEQ parton distribution functions
\cite{cteq4} to reproduce various hadronic spectra
\cite{hy}. Our attention in this paper is shifted from the
fragmentation region in hadronic collisions, where the $x$ dependence
is an issue, to the central region in nuclear collisions, where the
relative yields of particles produced are the focus.

Clearly, there is no way to study SE from first principles. Our
investigation here is  phenomenological. Our goal is very modest. It
is not possible to compute particle ratios from the parton model
alone. We shall use a large set of particle ratios as experimental
inputs to guide us in the determination of the SE factor. The effect
of Pauli blocking in the non-strange sector is included in the those
inputs, and are not amenable to first-principle calculations. Our
theoretical input is essentially the counting of quarks and
antiquarks in their partition into the various hadronic channels. In
that sense the physics involved is basically the same as in the
coalescence model
\cite{ab,zbc}, which is a simplified version of the recombination
model \cite{dh}. The emphases in Refs.\
\cite{ab,zbc} are  in the multiplicative aspect of the probabilities
of having quarks and antiquarks in the same region of phase space in
their formation of hadrons. That results in an undesirable feature of
$s$ and $\bar s$ imbalance in the linear version \cite{ab}, which is
not satisfactorily resolved in the nonlinear version \cite{zbc} by
the introduction of unknown factors. Our emphasis here is on the
partition of the $q, \bar q, s, \bar s$ quarks  into the hadronic
channels and on the enhancement of their populations from gluon
conversion. We shall not investigate the implications of the hadron
probabilities being products of the quark probabilities.

\section{Quark Counting}

We begin by drawing the boundary of our concern here. Since the
yields on multi-strange hyperons are low compared to
$K$ and
$\Lambda$, we shall in first approximation ignore the production of
$\Xi$ and $\Omega$, and aim at results with accuracies not better
than 90\%. With such simplification we can better exhibit the spirit
of our approach to the problem and make more transparent the issues
involved in SE. Improvements that include the $\Xi$ and $\Omega$
particles can be considered later. We shall also consider
isosymmetric dense medium at mid-rapidity so that we need not
distinguish $u$ and $d$ quarks. Proton and neutron will be equal in
number, as do $\pi^+$ and
$\pi^-$. The strange quarks $s$ and $\bar s$  are produced in equal
numbers, but $K^+$ and $\bar K^-$ will not be produced in equal
numbers because of associated production.

Let us use the following notation to denote the numbers of hadrons
and non-strange quarks, e.g., $N$ is the number of nucleons, and $q$
is the number of light quarks.
\begin{eqnarray} N&=&p+n,  \qquad\qquad \bar N=\bar p +\bar n,
\label{1} \\
\Pi&=&\pi^++\pi^-+\pi^0,  \label{2}\\
Y&=&\Lambda+\Sigma^0+\Sigma^++\Sigma^-,  \label{3}\\ K&=&K^++K^0,
\qquad\qquad \bar K=K^-+\bar K^0,  \label{4}
\\ q&=&u+d.
\label{5}
\end{eqnarray} Then there are linear relations among these numbers
based on counting the number of valence quarks in the various hadrons
\begin{eqnarray} 3N+\Pi+2Y+K&=&q,  \label{6}  \\ 3\bar N+\Pi+2\bar
Y+\bar K&=&\bar q,   \label{7}  \\ Y+\bar K&=& s
\label{8}  \\
\bar Y + K &=& \bar s.
\label{9}
\end{eqnarray} The right-hand sides of the above equations all refer
to the numbers of quarks after enhancement from gluon conversion. Let
$\kappa$ be the fraction of $s$ quarks that recombine with
non-strange antiquarks to form anti-kaons, and similarly
$\bar\kappa$ be the fraction of $\bar s$ to form kaons. That is, we
define
\begin{equation}
\bar K=\kappa\ s, \qquad\qquad  K=\bar\kappa\ \bar s.
\label{10}
\end{equation} Then on account of Eqs. (8) and (9), we have
\begin{equation} Y=(1-\kappa)\ s, \qquad\qquad \bar Y=(1-\bar\kappa)\
\bar s.
\label{11}
\end{equation} Define the hadronic ratios
\begin{equation} r=K/\bar K, \qquad\qquad  R=\bar Y/Y.
\label{12}
\end{equation} It then follows that
\begin{equation}
\kappa={1-R\over r-R}\ , \qquad\qquad  \bar\kappa=r\ \kappa,
\label{13}
\end{equation} where $s=\bar s$ has been used. For experimental
values of $r$ and $R$, we assume that
$K/\bar K\approx K^+/K^-$ and $\bar Y/Y\approx
\bar\Lambda/\Lambda$. For $Pb$-$Pb$ collisions at SPS the values are
\cite{gr,ea,ant}
\begin{equation} r=1.8, \qquad\qquad  R=0.13,  \label{14}
\end{equation} so we obtain
\begin{equation}
\kappa=0.52, \qquad\qquad  \bar\kappa=0.94.  \label{15}
\end{equation}

With these values of $\kappa$ and $\bar\kappa$ we can proceed to
consider the non-strange sector. Define
\begin{equation}
\rho=\bar N/N
 \label{16}
\end{equation} so that we can obtain from Eqs. (6) and (7)
\begin{eqnarray}
N&=&{1\over 3\,(1-\rho)}\left[q-\bar
q+3\,s\,(\kappa-\bar\kappa)\right],
\label{17}  \\
\Pi&=&{1\over 1-\rho}\left\{-\rho\,q+\bar q\nonumber\right.\\
&&\left.-s\,[(1+2\,\rho)
\kappa-(2+\rho)\,\bar\kappa+2(1-\rho)]\right\}.
\label{18}
\end{eqnarray}
 The particle ratios involving abundant strange and non-strange
hadrons are
\begin{eqnarray} {N\over K}&=&{1\over 1-\rho}\left[{q_v\over
3s\bar\kappa}+{1\over r}-1\right],  \label{19}  \\
{\Pi\over
K}&=&{1\over(1-\rho)\bar\kappa}\left[(1-\rho)\,{q\over s}-{q_v\over
s}\nonumber \right.\\
&&-\left.(1+2\rho)\,\kappa+(2+\rho)\,\bar\kappa-2\,(1-\rho)\right],
\label{20}
\end{eqnarray}
Where $q_v$ denotes the number of valence quarks, i.e., $q_v=q-\bar
q$. The RHS can be determined from parton distributions,
assuming that the parameters $\kappa,
\bar\kappa$ and $\rho$ are known from experiments. The LHS can be
related approximately to
$p/K^+$ and $\pi^+/K^+$:
\begin{equation} {p\over K^+}={N\over K}, \qquad\qquad {\pi^+\over
K^+}={2\,\Pi\over3\,K}.   \label{21}
\end{equation} The experimental values of these ratios at SPS are
\cite{ars,bl}
\begin{equation} {p\over K^+}=1.0, \qquad\qquad {\pi^+\over K^+}=4.76.
\label{22}
\end{equation}
  The values of $\rho$ is \cite{kan}
\begin{equation}
\rho=0.07.  \label{23}
\end{equation} With these experimental inputs there should be no
difficulty in satisfying Eq.\ (\ref{19}) and (\ref{20}) by varying
the quark numbers.

However, in our approach the quark numbers must fit into our scheme
of quark enhancement via gluon conversion.  Moreover, there is the
issue of what precisely is the central region where the experimental
numbers of the hadron ratios are measured. Clearly, the valence to
sea quark ratio depends on the region of small $x$ considered. The
experiments do not have a common and unique definition of the central
rapidity region. For our analysis in the following we define the
central rapidity region to correspond to a value
$x_0$, which depends on $\sqrt s$. In fact, when we make prediction
later on for RHIC energies we shall use the relation
\begin{equation} x_0=(s_0/s)^{1/2}.
\label{24}
\end{equation} For now at SPS we use $x_0$ as an adjustable
parameter. The important physics input is that for $x\le x_0$, which
is what Feynman called the ``wee" $x$ region, we assume that all
partons distributions are constant (consistent with the notion of
saturation) so that the quark number ratios in the wee region,
whatever the flavor, can be determined by computing the  ratios of the
corresponding quark distributions at $x=x_0$.

In the calculation of the quark distributions after gluon conversion,
we shall do it in a simplified way for the central region, different
from how it has been treated in the fragmentation region \cite{hy}.
The reasons are because firstly we need not distinguish $u$ and $d$
types quarks in the central region and secondly we need not track the
$x$ dependences. It is important to first refer all quark numbers to
those given by CTEQ at $Q^2=1$ GeV and
$x=x_0$. That is our baseline, from which we discuss enhancement in
the following. Now, CTEQ gives distributions, not number of partons.
For example, $u(x_0)$ is the probability of having a $u$ quark at
$x=x_0$. In the following we use $q_0$ to denote the number of $u$
and $d$ quarks in $x\le x_0$, and equate it to $u(x_0)+d(x_0)$,
multiplied by a  factor that is proportional to the relevant phase
space volume. Such a factor will cancel later upon taking the ratio
of quark numbers, so it will not appear explicitly in any of the
expressions for parton numbers below. Similarly, we use $s_0$ and
$g_0$ to denote the number of $s$ quark and gluons in the region $x\le
x_0$, but identified with $s(x_0)$ and $g(x_0)$, respectively, of the
CTEQ distributions.

\section{Gluon Conversion}

Before gluon conversion we have $q_v$ valence quarks,
$2\bar q_0$ non-strange sea quarks, and $s_0+\bar s_0$ strange
quarks. In the  case of hadronic collisions, it has been shown that
the inclusive distributions of produced hadrons can be reproduced
without any free parameters, if the gluons are completely converted
to  non-strange sea quarks before hadronization through recombination
\cite{hy}. Now, in the case of $AA$ collisions we must consider  the
conversion of gluons to strange quarks in addition to the non-strange
quarks because of Pauli blocking in the  light sector. We use
$\gamma$ to denote the fraction in the strange sector. That is, the
number of converted strange and non-strange quarks, labeled with
subscript $c$, are
\begin{equation} s_c=\gamma\,g_0, \qquad\qquad q_c=(1-\gamma)\,g_0,
\label{25}
\end{equation} with the corresponding antiquarks $\bar s_c$ and $\bar
q_c$ being equal in number, respectively. Thus after conversion we
have
\begin{eqnarray} &q=q_v+\bar q_0 + q_c,
\label{26} \\ &s=s_0 + s_c.
\label{27}
\end{eqnarray} The quark and gluon distributions at $x_0$ can be
either obtained from the graphs posted by CTEQ4LQ
\cite{cteq4}, or determined numerically from the analytic formulas
given in Ref.\ \cite{hy2}. We use the latter to fix $q_v, \bar q_0,
s_0$ and $g_0$ for every $x_0$, while
$q_c$ and $s_c$ depend on $\gamma$. Hence, we have two free
parameters,
$x_0$ and $\gamma$, to fit the data through the use of Eqs.
(\ref{19})-(\ref{23}).

From Eq.\ (\ref{19}) one gets $q_v/s=3.86$. Using that in (\ref{20})
yields
$\bar q/s=7.53$, whereupon one obtains
\begin{equation} {\bar q\over q}={1\over 1+q_v/\bar q}	= 0.66.
\label{28}
\end{equation} From (\ref{25}) and (\ref{27}) we have
\begin{eqnarray} s=s_0+\gamma\,g_0&=&q_v\,/\,3.86 ,
\label{29} \\
\bar q_0 + (1-\gamma)\,g_0&=&7.53\,s;  \label{30}
\end{eqnarray} together they give
\begin{equation}
\bar q_0+s_0+g_0=2.21\ q_v.   \label{31}
\end{equation} This is an equation that depends on CTEQ distributions
only, so we can solve for the value of $x_0$. The result then
determines also the values of $q_v, s_0$ and $g_0$, which, when used
in Eq.\ (\ref{29}), fix
$\gamma$. The process yields
\begin{eqnarray} x_0&=&0.135,  \label{32} \\
\gamma&=&0.08.   \label{33}
\end{eqnarray} The value of $x_0$ is reasonable, but the value of
$\gamma$ seems surprisingly low, since 8\% conversion from the gluons
seems insufficient to justify the notion of SE.

\section{Strangeness Enhancement}

To appreciate the value of $\gamma$ found above, let us examine the
quark distributions at $x_0$ before gluon conversion. Our solution of
Eq.\ (\ref{31}) gives
\begin{eqnarray}
x_0 \, q_v &=& 0.462, \quad  x_0\, \bar
q_0 = 0.118,  \nonumber \\
 x_0 \, s_0 &=& 0.052,\quad x_0 \, g_0 = 0.85.
\label{34}
\end{eqnarray}
Thus from Eq.\ (\ref{25}) we have
$x_0\,s_c=0.068$. Comparing $s_c$ with
$s_0$, we see that the strangeness enhancement factor $E_s$ at the
quark level is
\begin{equation}
E_s={s\over s_0}=1+{s_c\over s_0}=2.3.
\label{35}
\end{equation}
This indicates quite an appreciable amount of increase
of the strange quarks, qualitatively consistent with the hyperon
enhancement. The point is that there are so many gluons that an 8\%
conversion significantly enhances the strangeness content. The
remaining 92\% conversion to $q_c$ should be compared to 100\%
conversion in the case of hadronic collisions \cite{hy}. Let us call
the light quark population in the sea after 100\% conversion $\bar
q_1$, i.e.,
\begin{equation}
\bar q_1=\bar q_0+g_0.
\label{36}
\end{equation}
Then the change in the sea from $pp$ to $AA$
collisions can be characterized by the ratio $B$:
\begin{equation} B={\bar q\over\bar q_1}={\bar
q_0+(1-\gamma)\,g_0\over \bar q_0+g_0}=0.94.
\label{37}
\end{equation} This may be regarded as a numerical factor quantifying
Pauli blocking in the light quark sector. Note that it is less than
one by only a small amount, but enough to boost $E_s$ from one by
more than a factor of two.

The extension of this consideration to RHIC energies is
straightforward. We first use Eq.\ (\ref{24}) to determine
$s_0$ from the values of $x_0$ at SPS. Setting $\sqrt s = 17$ GeV, we
obtain
$\sqrt s_0=2.3$ GeV. Now, holding $s_0$ fixed, we have the
corresponding $x_0$ value (call it $x'_0$) at $\sqrt s=130$ GeV to be
\begin{equation} x'_0=0.0177.
\label{38}
\end{equation} The values of $q_v, \bar q_0, s_0$ and $g_0$ at $x'_0$
are (from CTEQ)
\begin{eqnarray}
 x'_0 q'_v = 0.149, \quad x'_0 \bar
q'_0 = 0.184, \nonumber \\
x'_0 s'_0 = 0.089, \quad x'_0 g'_0 = 1.229.
\label{39}
\end{eqnarray}
Note that $q'_v$ is much smaller than $q_v$, as
expected, so that
$\bar q'_0/q'_0=0.55$, even before gluon conversion. Thus we expect
antiparticle/particle ratios to be much closer to one.

As before, we need the experimental inputs at RHIC.  From Refs.\
\cite{hc,zx,ka} we have at $\sqrt s=130$ GeV
\begin{equation} r=1.136,\quad R=0.77,\quad \rho=0.64.
\label{40}
\end{equation} So we get from Eq.\ (\ref{13})
\begin{equation}
\kappa=0.628, \qquad\qquad \bar\kappa=0.713.   \label{41}
\end{equation} Assuming that $\gamma$ remains constant, we now can
calculate the quark ratios
\begin{equation} {q'_v\over s'}=0.8, \qquad\qquad {\bar q'\over
s'}=7.06,   \label{42}
\end{equation} which, when used in (\ref{19}) and (\ref{20}), enable
us to calculate the hadron ratios. As a consequence, we obtain
\begin{equation} {p\over K^+}=0.71, \qquad\qquad
{K^+\over\pi^+}=0.18.  \label{43}
\end{equation} The latter, compared to the value, 0.21, at SPS, is a
14\% decrease and agrees well with the data at RHIC \cite{jh}, which
shows
$K^+/\pi^+=0.176\pm0.004$. For the former we find indirect
confirmation from the following ratios reported by STAR
\cite{hc,star} for
$\sqrt s=130$ GeV:
\begin{eqnarray} {\bar p\over p}&=&0.65\pm0.07,
\qquad\qquad {\bar p\,\over
\pi^-}=0.08\pm0.01,   \nonumber  \\ {K^-\over
\pi^-}&=&0.149\pm0.02, \qquad\qquad {K^-\over K^+}=0.88\pm0.05.
\nonumber
\end{eqnarray} These numbers can be used to imply
\begin{equation} {p\over K^+}=0.73\pm0.1,   \label{44}
\end{equation} which agrees well with the calculated number in
(\ref{43}). These results give support to our assumption that
$\gamma$ is constant when the energy is increased and to our
procedure of treating the energy dependence.

The SE factor   becomes at RHIC ($\sqrt s=130$ GeV)
\begin{equation} E_s={s'\over s'_0}=1+{s'_c\over s'_0}=2.1.
\label{45}
\end{equation}
 Although the gluon density increases by 45\% as $x_0$ decreases to
$x'_0$, the $s$ quark density in the quiescent sea increases by even
more, so the net enhacement factor
$E_s$ decreases slightly. This small decrease is in agreement with
that of the statistical model \cite{rt}, although the physics is
totally different. The Pauli blocking factor becomes
\begin{equation} B=0.93,      \label{46}
\end{equation} which is essentially unchanged from Eq.\ (\ref{37}).

\section{Conclusion}

Since we have left out the multi-strange hyperons from our
consideration, we cannot expect the numbers calculated to be
accurate.  Moreover, the necessity to use such experimental inputs as
$r, R, \rho$ to determine $\kappa$ and $\gamma$ renders the approach
highly phenomenological, far from first principles. However, the
basic attributes of this line of study are to use the parton model
(and the distributions of CTEQ) as the basis for the investigation of
particle ratios in nuclear collisions at the quark level,  and to use
simple linear relations, Eqs.\ (\ref{6})-(\ref{9}), based on quark
counting as the only constraints among the strange and non-strange
hadrons. We have found consistency within this simple approach, and
can successfully describe the energy dependence. We have not assumed
thermal equilibrium, nor relied on the statistical model.  We have
also deliberately avoided treating mesons and baryons as products of
quark densities, as have been attempted in Refs.\ \cite{ab,zbc},
since they lead to either $s\ne \bar s$ or undetermined constants.

As we have stated at the outset, it is not our aim to predict
particle ratios. We have used the experimental values of the ratios
to lead us to the determination of the SE factor, $E_s$, and the
Pauli blocking factor, $B$, defined at the quark level. In so doing
we have gained some insight into how the enhancement mechanism works
through the process of gluon conversion. We have further learned that
a slight suppression of the conversion into the non-strange sector
gives rise to a substantial increase in the strange sector. Such a
small change from hadronic to nuclear collisions makes strangeness
enhancement an unreliable signature for the formation of  quark-gluon
plasma.

\section*{Acknowledgment}

We are grateful to A.\ Capella, D.\ Kharzeev, R.\ Lietava and A.\
Tounsi for helpful discussion and communication. This work was
supported, in part,  by the U.\ S.\ Department of Energy under Grant
No. DE-FG03-96ER40972.

\end{document}